\documentclass[aps,pre,twocolumn]{revtex4-1}

\usepackage{epsf}
\usepackage{latexsym}
\usepackage{epsfig}
\usepackage{amsmath}
\usepackage{amsfonts}
\usepackage{amssymb}
\usepackage{graphicx}
\usepackage{color}
\usepackage[usenames,dvipsnames]{xcolor}

\newcommand{\prs}[1]{{\left(#1\right)}}
\newcommand{\col}[1]{{\left[#1\right]}}
\newcommand{\chs}[1]{{\left\{#1\right\}}}

\newcommand{\vc}[1]{{\boldsymbol #1}}

\newcommand{\sgn}{{\mbox{sgn}}}

\begin{document}

\title{A Model for Emergence of Multiple Anti-Microbial Resistance in a Petri Torus}
\author{Alamino, R.C.}
\affiliation{Aston University, Birmingham B4 7ET, UK}

\begin{abstract}
This work introduces a new statistical physics lattice model of bacteria interacting with anti-microbial drugs that can reproduce qualitative features of resistance emergence and whose model parameters and outputs can be measured with controlled \textit{in vitro} experiments. The lattice is inhabited by agents modeled by Ising perceptrons. The results show the advantage of mixing drugs among the population compared to other treatment protocols.
\end{abstract}

\maketitle

\section{Introduction}

Anti-microbial resistance (AMR), the resistance of infectious micro-organisms to therapeutic drugs, has been dramatically increasing during the last decades, becoming a threat as dangerous as climate change \cite{Woodford09, Andersson10} and potentially more urgent. With decreasing rates of anti-microbial drug (AMD) development \cite{Charles04}, we risk returning to a defenseless era against infections. The issue is so important that the World Health Organization has proposed a global action plan to address it \cite{WHO15}. 

The focus of most studies lies on bacteria, which can develop AMR in several ways. In addition to mutations in their main DNA and the possibility of horizontal gene transfer (HGT) through the intermediate action of phages (bacteria-infecting viruses), they also possess circular DNA fragments, called \emph{plasmids}, which frequently encode resistance genes \cite{Ng10}. Plasmids can also mutate like the main DNA and they largely contribute to HGT \cite{Millan15, Baltrus13, Andersson10} by being exchanged between bacteria of different species or freed into the environment for instance, upon cell death, once the cell's membrane breaks down. This versatility makes plasmids powerful tools for AMR acquisition.

There are several attempted methods to deal with resistance, although none of them has been enough to alleviate the problem satisfactorily. One common practice is simply to avoid using of some AMD for which resistance has emerged for a (potentially long) amount of time. Although resistance seems to decrease in general with this protocol, its efficiency has been disputed \cite{Barbosa00}. Evidence shows that the reversal rate is slow \cite{Austin99, Tan11} and, even after \emph{some} reversal is observed, AMR never goes back to original levels.  

Other common treatment protocols make simultaneous use of more than one AMD \cite{Bonhoeffer97}, opening the doors for the emergence of multiple resistance. The most popular are the two-drug protocols called \emph{combining}, \emph{mixing} and \emph{cycling}. Combining AMDs amounts to administering the two drugs simultaneously to all patients. Mixing means that each drug is given separately to a certain fraction of the patients, who keep taking the same drugh during the whole treatment period. Finally, cycling entails the periodic alternation between drugs, each one given to the whole group at the same time.

There is a large literature on modeling AMR using differential equations \cite{Bonhoeffer97, Alexander07, DAgata08, Obolski12, Ternent15}. From the perspective of statistical physics, these are effective models, i.e., averages over large populations. They have been used to predict and analyze different aspects of AMR. However, as it is common with effective models, they comprise a large number of adjustable parameters as many of them have to be included \textit{ad hoc} to allow for a better adjusting of observed features.

In microscopic models, on the other hand, the number of adjustable parameters is usually reduced as they should emerge from microscopic quantities of the system. While effective models can be advantageous from the practical point, once their solution has usually a lower computational cost, microscopic models help in understanding basic principles behind the phenomenon, leading possibly to a better control and refinements or corrections to the effective models.   

AMR and evolution of cell cultures have been addressed in different branches of theoretical physics before, but most models were again effective ones \cite{Patra14, Lambert15, Bittihn17}. The use of microscopic models has been far less common \cite{Hermsen10, Greulich12}. Analytical results from related statistical physics evolutionary models also exist, but AMR is not specifically addressed \cite{Tikhonov17}. 

Given the importance of the subject, it is surprising that the powerful methods of statistical physics have not been more widely applied to the AMR problem. This paper contributes to fill this gap by introducing a new microscopic model of agents in a lattice. Using fewer parameters than models relying on differential equations, it is capable of reproducing qualitatively main features of AMR. It is clear that simplified models like the present one cannot reproduce perfectly all the subtleties of the biological processes involved in AMR. In this first approximation though, the aim of this work is to reproduce mainly qualitative aspects and understand the fundamental mechanisms at work. Quantitative agreement with experiments requires more thorough studies and model refinement.    

In section \ref{section:AB} we introduce the lattice model representing an artificial bacteria culture in a Petri torus. The model has only three parameters representing a genotype-phenotype map from the cell's plasmid to their response to AMDs. The introduced model is used to analyze the effect of the single drug protocol in section \ref{section:SR}, where it is shown that it can reproduce the memory effect we mentioned above. Section \ref{section:DR} shows the results for double-drug protocol in which we single out the mixing protocol as the best option to slow down the emergence of resistance. Finally, further discussions and conclusions are presented in section \ref{section:Conc}.
 
\section{Artificial Bacteria} 
\label{section:AB}

The model presented here corresponds to a scenario in which a bacterial culture in a Petri dish is subjected to different treatments with different AMDs. The evolution of the culture in time is then followed in order to provide information about how each treatment affects the dynamical aspects of AMR emergence in the culture.

For convenience, the Petri dish is represented in a simplified way by a square $N\times N$ lattice with periodic boundary conditions (PBC) in both directions, which is going to be called a \emph{Petri torus} due to the resulting topology. To each site $(i,j)$, $i,j=1,...,N$, a spin variable $\sigma_{ij}$ which is +1 if the site is occupied by a cell and -1 if it is empty. PBC imply $\sigma_{i+N,j}=\sigma_{i,j+N}=\sigma_{ij}$.     

The main bacterial life cycle, excluding the action of the AMDs, is modeled by two parameters. At each time step $t$, cells reproduce by choosing one of its four neighboring sites with the same probability. If the site is empty, then there is a probability $r$ for the cell to divide with the newborn bacteria occupying the new site. A \emph{natural death probability} $d$ for each living cell at time $t$ is used to model all \emph{non-drug} related processes like other adverse environmental conditions, the natural life-cycle of the cell and a patient's immune system. 

Bacteria are modeled as agents living in the Petri torus subjected to external stimuli provided by local concentrations of the different AMDs. The drugs are administered according to each specific treatment protocol to be analyzed. Pharmaceutical companies usually measure the efficiency of particular AMDs by their \emph{Minimum Inhibitory Concentration (MIC)}, the drug concentration above which bacterial growth stops on average \cite{Davey15}. The MIC is convenient in clinical trials as it avoids the difficulties in isolating the effects of patients' immune system, but it contains no information about pharmacodynamic properties of the drug (how bacterial growth changes with variations in drug concentrations). It has been proposed that a better proxy is given by the \emph{Minimum Bactericidal Concentration (MBC)}, the concentration that kills at least 99.9\% of the bacteria on average \cite{French06}. Here, a variation of the MBC is proposed and used: the concentration below which no cell dies. This definition is much simpler to implement in a probabilistic microscopic model and, because the studied scenario comprises \emph{in vitro} cultures, it can be actually measured with controlled experiments.  

Finally, it is assumed for simplicity that each cell has one single plasmid encoding its AMD response, which is given by the \emph{total AMD death probability}  $q^\mu_{ij}$, the probability that the cell occupying the site $(i,j)$ dies if exposed to the \emph{local} concentration $c_{ij}^\mu$ of the $\mu$-th AMD, $\mu\in\chs{1,...,M}$, where $M$ is total number of different available AMDs. This probability is modeled by 
\begin{equation}
  q_{ij}^\mu = \Theta\prs{\Delta^\mu_{ij}} p_{ij}^\mu \prs{1-e^{-\lambda_{ij}^\mu\Delta^\mu_{ij}}}.
\end{equation} 
where $\Delta^\mu_{ij} \equiv c_{ij}^\mu-\bar{c}_{ij}^\mu$.

There are only three parameters in the above expression, what is already a huge simplification if compared to the models based on differential equations. The \emph{MBC} $\bar{c}_{ij}^\mu\in\left[0,\infty\right)$, enters in the model through the Heaviside step function $\Theta(x)$, which is zero if $x<0$ and 1 otherwise. Its objective is to account for physical and chemical mechanisms having threshold behaviors as, for instance, chemical pumps which can become saturated or membranes whose thickness up to a certain point can prevent the AMDs from entering the cell interior. The \emph{maximum death probability} $p_{ij}^\mu\in\col{0,1}$ is a scaling that sets the maximum damage a certain AMD can inflict to the cell. Finally, the \emph{sensitivity} $\lambda_{ij}^\mu\in\left[0,\infty\right)$ regulates the increase in cell death with AMD concentration and is related to the actual \emph{toxicity} of it. All three parameters can, in principle, be obtained from actual designed experiments by measuring changes in bacterial populations.

The genotype of each plasmid will be encoded by a spin chain $\vc{\pi}_{ij}\in\chs{\pm1}^D$, where $D$ is an integer. The crucial point in epigenetics is to find an appropriate genotype-phenotype map into the model parameters \cite{Stadler06, Cortini16}. The map should allow for different genotypes to generate the similar phenotypes and also for \emph{learning} and \emph{forgetting}. Learning is required for adaptation in the form of AMR acquisition and forgetting allows acquired adaptations to fade away, as would be the case for reversal of AMR. One of the most studied statistical physics models of biological phenomena with these characteristics is the \emph{Ising perceptron} \cite{Rosenblatt58, Engel01}, designed to model neuronal responses, with the neuron's synapses and stimuli both encoded by spin chains. 

The perceptron is one of the simplest known learning machines and it is characterized by a function taking a multidimensional vector into a number, called the \emph{activation function} for biological reasons. This is usually a general function of the scalar product of its \emph{synaptic vector}, which is a multidimensional parameter encoding the information learned by the perceptron, and the \emph{input vector}, a vector with the same dimensions as the synaptic vector and which encodes the stimuli provided by the environment to which the perceptron reacts.  

In this work, the input vectors correspond to the binary encodings of each AMD into a spin chain of dimension 3D, $\vc{A}^\mu=\prs{\vc{\alpha}^\mu,\vc{\beta}^\mu,\vc{\gamma}^\mu}\in\chs{\pm1}^{3D}$. The three model parameters are then given by
\begin{align}
  \bar{c}_{ij}^\mu &= \frac{1+\phi_{ij}^\mu}{1-\phi_{ij}^\mu}, \qquad \phi_{ij}^\mu = \frac{\vc{\alpha}^\mu\cdot\vc{\pi}_{ij}}{D} ,\\ 
  \lambda_{ij}^\mu  &= \frac{1+\xi_{ij}^\mu}{1-\xi_{ij}^\mu}, \qquad \xi_{ij}^\mu = \frac{\vc{\beta}^\mu\cdot\vc{\pi}_{ij}}{D} ,\\ 
  p_{ij}^\mu       &= \frac{1+\omega_{ij}^\mu}2, \qquad \omega_{ij}^\mu = \frac{\vc{\gamma}^\mu\cdot\vc{\pi}_{ij}}{D}, 
\end{align}
which are chosen for being the simplest mappings into the relevant intervals.
 
Although simple perceptrons cannot approximate general functions, it has been shown that adding one extra \emph{layer}, corresponding to another set of perceptrons realising an intermediate processing of information between the stimuli and the final response, turns them into universal approximators \cite{Cybenko89}. Variations with several layers, known as deep neural networks, have been successfully used in machine learning applications and recently provided a solution for the long sought problem of creating a computer algorithm capable of playing Go on a level comparable to human masters \cite{Silver16}. 

The dimension of the AMD vector was chosen to be three times larger than the plasmid's one to allow mutations in the latter to simultaneously affect all three model parameters. In nature, each protein usually participates in more than one metabolic process simultaneously. As a consequence, each single mutation might affect more than one of them. 

One should notice that, although $p$, $\bar{c}$ and $\lambda$ are being called the \emph{parameters} of the model, they are actually not numeric adjustable parameters. Once the genotype and the AMD are chosen, they are fully determined. More precisely, the only freedom comes from the choice of the activation functions leading to them. One might call the latter the \emph{functional parameters} of our model, as it remains open the possibility of choosing more complicated genotype-phenotype maps which might lead to better agreement between observation and theory if needed.

\section{Single Resistance} 
\label{section:SR}

In order to obtain the dynamical behaviour of the AMD death probability $q$ for one single drug, the model is simulated in discrete time $t$. Each initial Petri torus occupation is set by putting a cell in each site with probability 1/2. All initial cells carry the same all-ones plasmid $\vc{\pi} = \vc{1}\equiv (1,1,...,1)$. To isolate the effect of the AMD, we will use $r=1$ and $d=0$, i.e., cells always reproduce if there is space and do not die unless killed by the AMD. With these two parameters fixed to these values, the dynamics then follows two steps at each $t$:

\noindent (1) \emph{Reproduction:} all living cells are drawn once and only once with the same probability. Then, one of the four neighbors of that cell is chosen with probability 1/4 and, if the corresponding site is empty, the cell generates a child cell on it. The child cell has a probability $m$ of mutation, where one single coordinate of its plasmid, chosen at random with equal probability, is flipped.

\noindent (2) \emph{AMD Death:} in a uniformly random order, one checks whether each cell dies according to the probability $q_{ij}^\mu$ for each AMD present locally.

Each run of the simulation consists of $T$ time steps and the information recorded is a double average of the death probability $q^\mu_{ij}$ - the average over all living cells and the \emph{quenched average} over initial configurations. 

According to our model, bacteria acquire resistance by (i) increasing the MBC, (ii) decreasing the maximum death probability, (iii) decreasing their sensitivity to the drug or any combination of these. Although the same mechanisms can contribute to different parameters, the larger the plasmid dimension $D$ is, the more uncorrelated we expect them to be. 

\begin{figure}
	\centering
	\includegraphics[width=7cm]{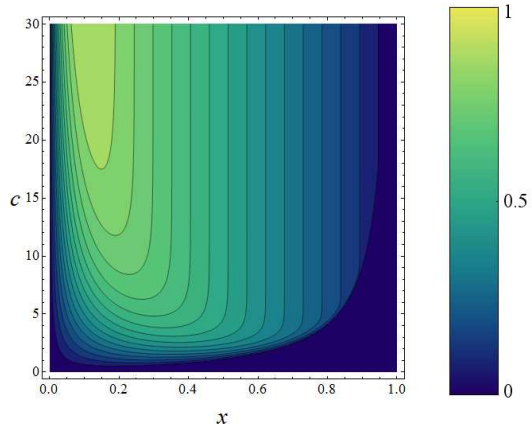}
	\caption{\textbf{Fitness landscape.} The plot shows the fitness landscape for the AMD $(-\vc{1},-\vc{1},\vc{1})$ as a contour plot representing the total death probability $q$ as a function of the AMD concentration $c$ and the fraction $x$ of -1 coordinates in the plasmid. The landscape shows a global maximum of $q$ at the centre extending to the top of the graph surrounded by decreasing profiles to both side. While higher concentrations will eventually kill all cells leading to $q=1$, one can see that by changing $x$ in any direction decreases $q$ and, therefore, increases resistance.}
	\label{figure:fitness}
\end{figure}

It is clear that for large dimensions $D$, which is the case in practice, the fitness landscape will be highly AMD dependent. It is reasonable to expect it to b also very complex, with several local minima and maxima which cannot be easily studied only with analytical methods. In order to get some insight into these landscapes, we start our analysis by illustrating how one can use the model to extract useful information in simple cases. One such case is the choice $\alpha_i=\beta_i=-1$ and $\gamma_i=1$, $i=1,...,D$, in which the parameters can be expressed as
\begin{equation}
\bar{c}=\lambda=\frac{x}{1-x}, \qquad \text{and} \qquad p=1-x,
\end{equation}
where $x=n/D$ and $n$ is the fraction of -1 coordinates in the plasmid. When $D\rightarrow\infty$, $x$ becomes a continuous real variable in the interval $[0,1]$. In this case it becomes very simple to plot the fitness landscape as a contour plot of $q$ as a function of $x$ and the AMD concentration $c$ as given in fig. \ref{figure:fitness}. 

In this particular case, $\bar{c}$ and $\lambda$ are equal by design, which will usually not be the case in practice. Still, qualitatively this is an important situation where they both contribute in different ways for the emergence of AMR. While a higher MBC improves resistance, a higher sensitivity decreases it. This is reflected in the fitness landscape by the maximum of $q$ surrounded by descending profiles both to the left and right. Because this landscape has a only one global maximum, we can more clearly see that resistance will eventually emerge as we move away from it in the $x$ direction, which is always observed in simulations.  

\begin{figure}
  \centering
  \includegraphics[width=8.5cm]{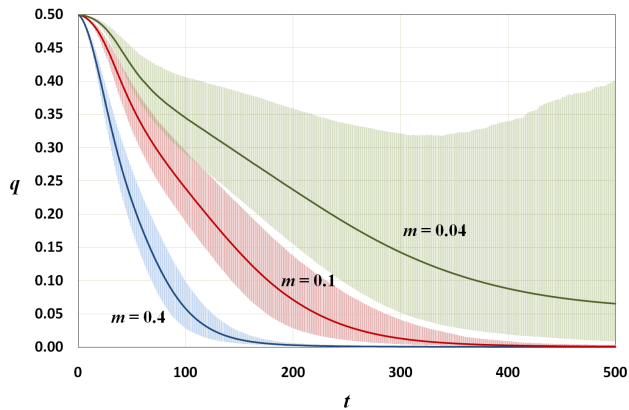}
  \caption{\emph{Average death probability} $q$ as a function of time $t$ for three different mutation rates (labels near each corresponding line). Shaded areas around the curves represent variances.}
  \label{figure:RAMD}
\end{figure}

For the rest of this work, we use AMDs with $D=70$ where $\vc{\alpha}$ and $\vc{\beta}$ are generated randomly and uniformly. Without loss of generality, we set $\gamma_i=1$ to tune the initial value of $p$ to 1. Fig. \ref{figure:RAMD} shows the average death probability $q$ for mutation rates $m=0.4, 0.1, 0.04$ in a lattice of size $N=100$. Results are averaged over $K=100$ realizations of the same AMD, differing by the initial site occupation, and over $N_a=10$ different randomly generated AMDs. Averaging over AMDs drastically increases the deviations from the mean curves as the response to each might be very different for the same plasmid. The local concentrations of AMD are kept constant during simulations, fine-tuned to give an initial $q=0.5$. As expected, larger mutation rates allow for faster adaptation (smaller $q$). The shadowed areas around the mean curves represent deviations from mean plots and, as discussed before, it is clear the dramatic increase in their amplitude as the mutation rate is decreased. 

Because the death probability is restricted to the interval $\in[0,1]$, deviations above and below the mean are calculated separately. Let $\bar{q}_{kl}$ be the average death probability for the $k$-th run of the $l$-th AMD. Deviations above and below the mean are respectively
\begin{align}
  \sigma^2_+ &= \frac1{n_+} \sum_{k,l} \Theta\prs{\Delta q_{kl}} \prs{ \Delta q_{kl}}^2, \\
  \sigma^2_- &= \frac1{n_-} \sum_{k,l} \col{1-\Theta\prs{\Delta q_{kl}}} \prs{ \Delta q_{kl}}^2,  
\end{align}
where $\Delta q_{kl} = q-\bar{q}_{kl},$ $n_+$ and $n_-$ are the number of cases in which the variation is respectively non-negative or negative.

Deviations become larger for smaller mutation rates $m$ because different AMDs lead to very different rates of emergence. With higher $m$, cells probe larger areas of the fitness landscape before dying, resulting in faster growth of AMR. Higher initial concentrations require greater $m$'s for adaptation to win over extinction. The exact relationship between this threshold and the other parameters of the model requires a better characterization of the phase diagram, which is under study. Realistically, higher concentrations become toxic also to the patients and, although efficient to kill bacteria, might result in potentially fatal side effects.

\begin{figure}
  \centering
  \includegraphics[width=8.5cm]{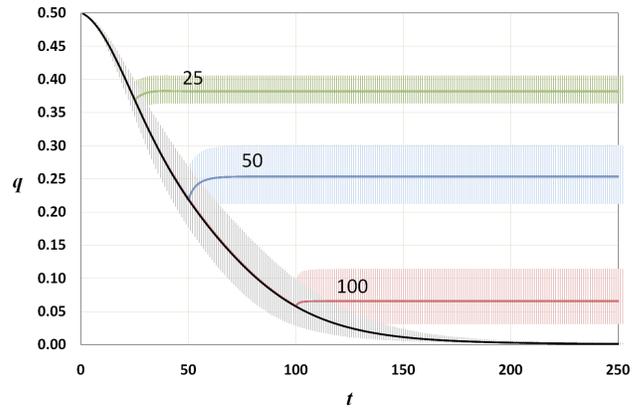}
  \caption{\emph{Average death probability} $q$ as a function of time $t$ for $m=0.4$. Each treatment is stopped at a different time (curve labels). The bottom line shows a continuing treatment. Shaded areas around curves represent deviations.}
  \label{figure:memory}
\end{figure}

Fig. \ref{figure:memory} reproduces the observed memory effect that stopping the treatment does not restore resistance to its levels before its beginning. The same parameters as before are used, but only for $m=0.4$, averaging over 10 AMDs. Stopping times are written next to the corresponding curves, the bottom one representing a continuing treatment.

\section{Double Resistance}
\label{section:DR}

While it seems logical to use several AMDs simultaneously, as the probability of being resistant to all of them is surely smaller than to a single one, this creates a multiple selection pressure that might lead to the emergence of resistance to all of them. To study this, we simulate the currently most common two-drug protocols: (i) combining, (ii) mixing and (iii) cycling.

\begin{figure}
	\centering
	\includegraphics[width=8.5cm]{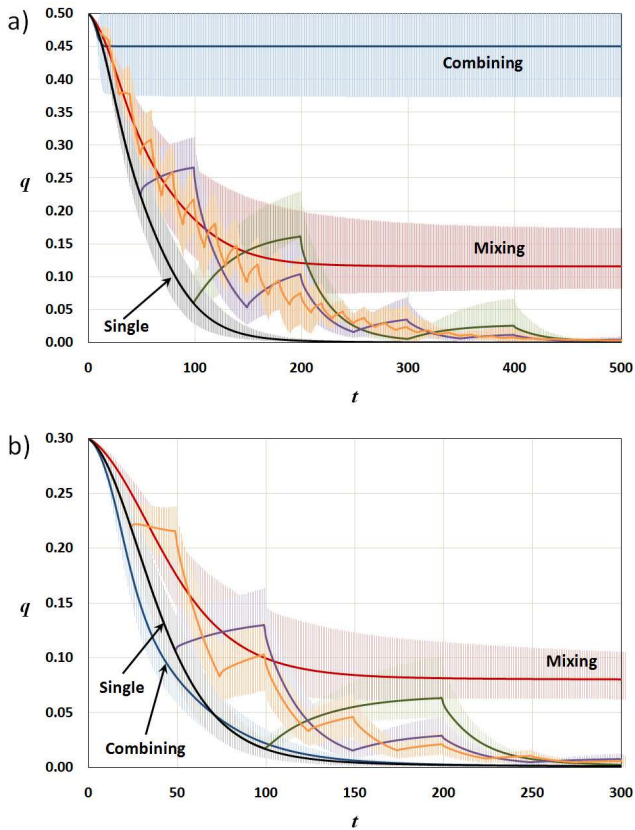}
	\caption{\emph{Average death probability} $q$ as a function of time $t$ for different protocols: single AMD, combining, mixing (indicated by labels) and cycling (spiked curves) with periods 10, 50 and 100. In multi-drug protocols, only one of the AMDs is shown. Shaded areas around curves indicate deviations. Simulations were run for initial (a) $q=0.5$ and (b) $q=0.3$ and a mutation rate $m=0.4$ in a lattice with 10000 sites.}
	\label{figure:double05}
\end{figure}

Fig. \ref{figure:double05} compares the rate of emergence of resistance when only one AMD (Single) is used against that of the three two-drug protocols. The curves are averages over 100 random initial occupations of the lattice and 10 randomly generated pairs of AMDs. Each AMD is generated independently, its coordinates being $\pm1$ with equal probabilities. The plot shows only the first AMD of each pair. The results for the second vary slightly quantitatively, but the qualitative behavior is similar.

All protocols were applied using the same initial death probability, $q=0.5$ for the upper plot and $q=0.3$ for the bottom one. $N=100$ and $m=0.4$ for both. Again, shaded areas around curves represent deviations above and below mean values.

For the combining protocol, both AMDs are present at each site simultaneously. Concentrations are site independent and do not vary with time. For the mixing protocol, only one AMD is present at each site. For every site, we choose one AMD of the pair with probability 1/2 and keep its concentration constant in time. All sites with the same AMD have the same drug concentration. For cycling, the simulation starts with the first AMD in each pair present in all sites at the same concentration. After each cycling period $C$, the AMD is changed by the other one in the same pair in all sites. Results are presented for three different periods, $C=10, 50$ and 100.  
  
For the chosen parameters, combining shows very little emergence when $q=0.5$. However, this is mainly a result of the double concentration to which cells are exposed. Because the initial $q$ is relatively high for both AMDs, cells die faster than they can adapt. Mixing seems to be the next best, with an apparent asymptotic stop of resistance emergence. Although the cycling protocols do better than the single drug case, they seem to be unable to avoid the emergence of total resistance, which in this case occurs for \emph{both} AMDs. In particular, this is independent of the cycling period, whose only effect is to create larger oscillations about an average line roughly coinciding with the smaller cycle. For cycling, we see again the memory effect discussed in the introduction: when one AMD is not being used, resistance decreases but not enough to bring a complete recovery to initial levels. It is not clear if there is any influence of the second AMD in the recovery, an interesting question for further studies.

For the plot in fig. \ref{figure:double05}b, simulations start with $q=0.3$ by appropriately adjusting the AMD concentrations. Differently from the previous case, combining becomes the worst of the multi-drug protocols, \emph{worse} even than the single-drug one after some time. This change is explained by noticing that with $q=0.3$, cells live long enough to allow AMR to appear through mutations. The data indicates again that, while other protocols result in the emergence of an almost total resistance, mixing saturates at a lower level for the given parameters. 

\begin{figure}
	\centering
	\includegraphics[width=8.5cm]{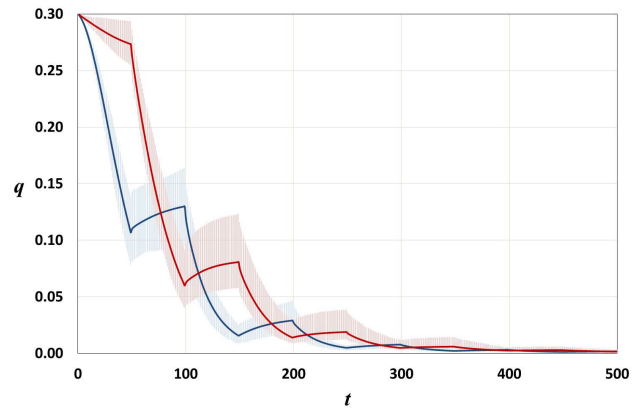}
	\caption{\emph{Average death probability} $q$ as a function of time $t$ for cycling with period $C=50$ for both AMDs in the pairs for the same parameters as in fig. \ref{figure:double05}b.}
	\label{figure:cyccomp}
\end{figure}

Fig. \ref{figure:cyccomp} shows the same graph for the two drugs of the pair for a cycling protocol with period $C=50$. The first AMD in the pair is represented by the line which drops faster in the initial period.

We measured the joint pharmacodynamics of a certain AMD pair by the average relative sign of their variations of $q$
\begin{equation}
  \rho(t) = \frac1{K N_a} \sum_{k,l} \sgn\prs{\Delta\bar{q}^1_{kl}(t)\Delta\bar{q}^2_{kl}(t)},
\end{equation}
where $\sgn\,x = x/|x|$ if $x\neq0$ and 0 otherwise, $\Delta\bar{q}^\mu_{kl}(t)=\bar{q}^\mu_{kl}(t)-\bar{q}^\mu_{kl}(t-1)$ and $\bar{q}^\mu_{kl}(t)$ is the death probability averaged only over lattice sites for the $\mu$-th AMD in the $l$-th pair during the $k$-th simulation run at time step $t$.

\begin{figure}
  \centering
  \includegraphics[width=8.5cm]{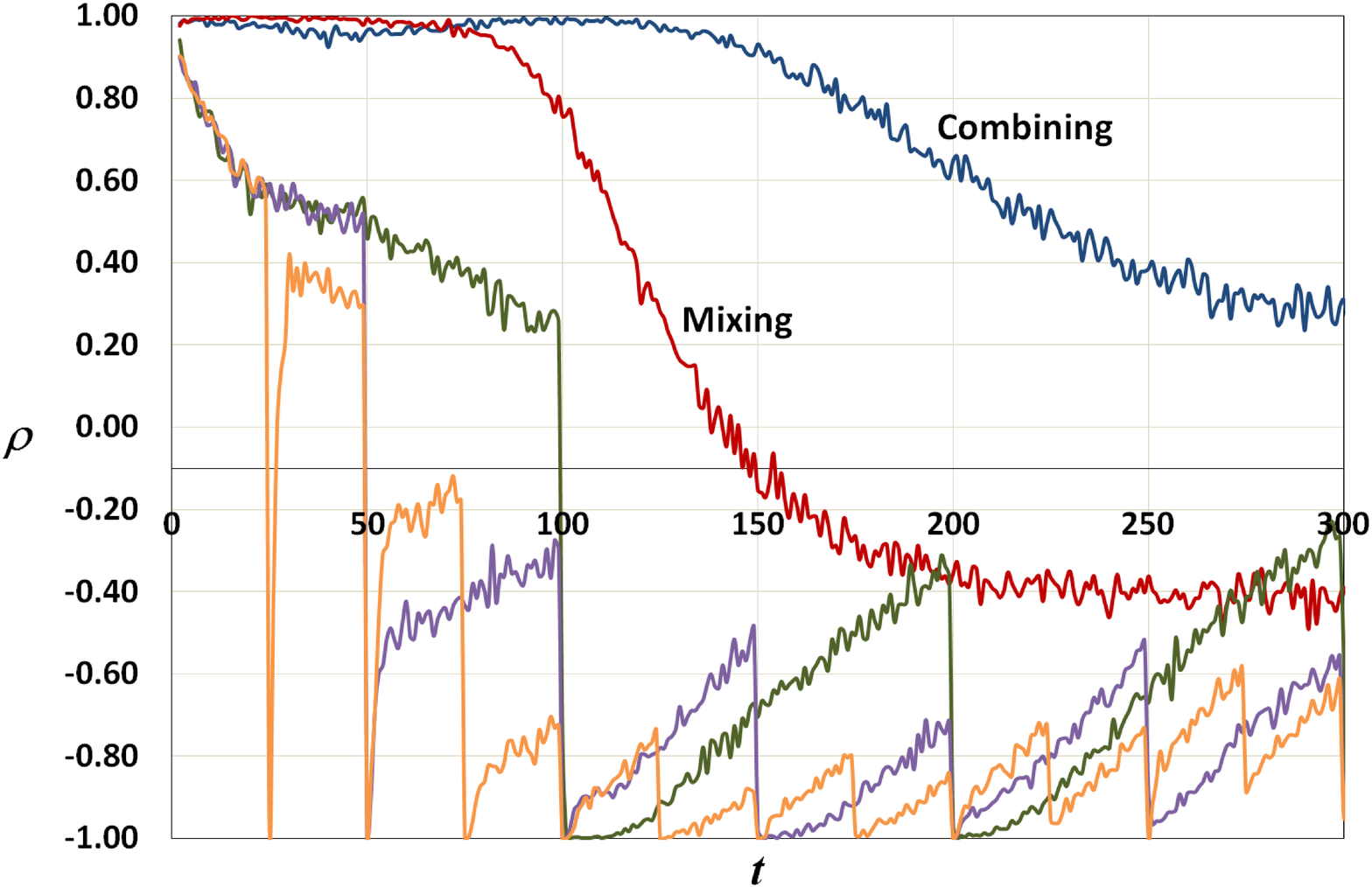}
  \caption{Average value of the relative sign of the variations in $q$ between the two AMDs in each pair. Results for combining and mixing are labeled in the plot. Spiked curves represent cycling with three different periods -- 25, 50 and 100 steps -- easy to identify due to the periodic behaviors. The simulation parameters are the same as in fig. \ref{figure:double05}b.}
  \label{figure:correl}
\end{figure}

This quantity, plotted in fig. \ref{figure:correl} for each two-drug protocol (using the same parameters as in fig. \ref{figure:double05}b), shows how often on average the cells have the same adaptive trend towards resistance to the two drugs in the pair. The value +1 means that both $q$ change in the same direction, i.e., bacteria are either becoming resistant or more susceptible to both AMDs simultaneously. Conversely, -1 means that their adaptive trend is opposite, they are gaining resistant to one of them and losing to the other.

The results indicate that when both AMDs are simultaneously present, resistance changes in the same direction for longer than in cycling, but this is not enough for the latter to outperform mixing. Mixing remains superior because, when reproducing, cells will half of the time spread to a site with a different AMD, to which they are less resistant. Although this seems similar to cycling, the latter has the whole population subjected to the same AMD at every time step. 

\begin{figure}
  \centering
  \includegraphics[width=8.5cm]{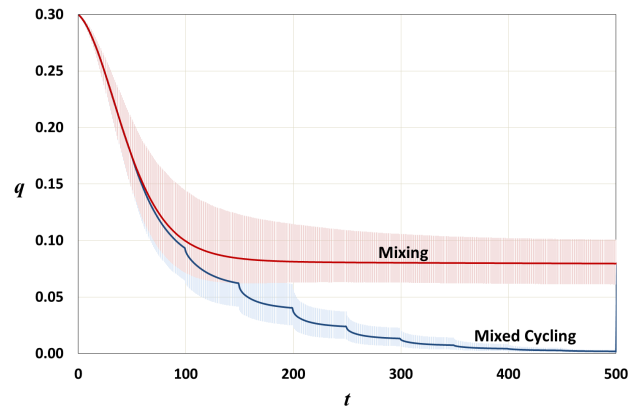}
  \caption{Comparison between \emph{mixed cycling} and (plain) mixing with the same parameters as in fig. \ref{figure:double05}b and cycling period of 50 steps. Mixed cycling is not enough to outperform mixing with AMR emergence rates surprisingly being increased for each cycling period instead of decreasing as for pure cycling.}
  \label{figure:mixcyc}
\end{figure}

We now analyse a new protocol, \emph{mixed cycling}, which in principle should combine the advantages of mixing and cycling: we distribute two AMDs uniformly randomly to the population and then cycle the resulting configuration. Fig. \ref{figure:mixcyc} shows a comparison between this protocol and mixing.

The plot, in which the same parameters as in fig. \ref{figure:double05}b were used and where the cycling period was chosen to be $C=50$, shows the results for the first AMD in the pair. One can see that mixed cycling ends up performing \emph{worse} than mixing. Unexpectedly, instead of observing a reversal in resistance at each cycling period, the change in AMD has exactly the opposite effect: it increases the AMR emergence rate. The mechanism at work here is subtle, but at the same time very simple. With pure cycling, each period sees adaptation to only one AMD, while the other loses it. Resistance is reversed because no cell is under pressure to adapt to the absent AMD.

On the other hand, in every cycling period of mixed cycling, half of the cells are subjected to one AMD and half to the other. This implies that adaptation to both, although at a smaller rate, is always taking place. Still, that does not explain the whole story as, in the mixing protocol, both AMDs are also present all the time during the runs. The difference is that each cell which has already adapted in mixing, will not suffer any different pressure as the AMD at that site never changes. On the other hand, with the cycling, it is force to re-adapt. That, instead of decreasing AMR, works in its favor. The details of this mechanism need to be studied more carefully and is the subject of our current research.

\section{Conclusions}
\label{section:Conc}

The model presented here has many advantages complementary to usual ones based on differential equations. Being a microscopic statistical model, there is larger control over mechanisms for AMR emergence. It also benefits from clarity and versatility while, at the same time, it can reproduce qualitatively actual features of both single and double resistance to arbitrary AMDs using very few parameters. 

Additional mechanisms, like exchange or acquisition of plasmids, are easy to include in the model. One exciting possibility is to use machine learning algorithms to encode the structure of actual AMDs and study them. The plasmid genotype can be directly translated to binary code and the macroscopic parameters of the model can be obtained from experiments. Although perceptrons are too simple to approximate general genotype-phenotype maps, it was proven \cite{Cybenko89} that more complex networks, as deep networks \cite{Silver16}, are universal approximators and can become powerful tools in the search of real new AMDs and evaluation of resistance scenarios.  

By applying the model to the most commonly used treatment protocols, we showed that the model (i) is capable of producing single and double resistance and (ii) can reproduce the memory effects observed in practice which indicates that reversal rates in the absence of AMD are inefficient to recover initial levels of resistance.

The simulations that have been carried out indicate that the mixing protocol is the most efficient of the protocols. The reason seems to be deceptively simple: while in all other protocols the whole population is subjected to the same AMD for some period of time, resistance grows faster during these times as all bacteria evolve in the same direction. With mixing, half of the population adapts mainly to one AMD and the other half to another. In addition, whenever bacteria spread, they have a chance of meeting a different AMD than the one in its original site, which changes the adaptation trend. Evidence in favour of this explanation comes from the proposed mixed cycling protocol. One could expect that this would benefit from the advantages of both mixing and cycling. In particular, the reversal in resistance observed at each cycling period. However, because of the mixing in the population, when the cycling period is reached, half of the bacteria are still subjected to the same AMD and, although slower, adaptation continues leading to resistance. On the other hand, contrary to mixing, all bacteria are exposed to both AMD, although in different periods. Because of this, cells at a definite site have contact to both AMDs instead of only one as in mixing, which leads to higher average adaptation to both.  

Other statistical physics issues need further analysis as, for instance, finite size effects and the existence or not of true phase transitions to resistant phases. HGT might be an important issue as it is known to change the critical transition point for speciation \cite{Park07} and should also affect significantly AMR emergence.

Many qualitative aspects of the model are fairly general and actual experiments are fundamental in testing its predictions as well as identifying aspects in which a more sophisticated modeling is needed in order to allow for a closer approximation of practical situations.

\section*{Acknowledgements}
I would like to thank Dr J. Neirotti, Dr M.Stich, Dr M.Chli and Dr A. Cheong for their suggestions and discussions. 


\bibliographystyle{apsrev4-1}
\bibliography{amr}

\begin{thebibliography}{30}%
\makeatletter
\providecommand \@ifxundefined [1]{%
 \@ifx{#1\undefined}
}%
\providecommand \@ifnum [1]{%
 \ifnum #1\expandafter \@firstoftwo
 \else \expandafter \@secondoftwo
 \fi
}%
\providecommand \@ifx [1]{%
 \ifx #1\expandafter \@firstoftwo
 \else \expandafter \@secondoftwo
 \fi
}%
\providecommand \natexlab [1]{#1}%
\providecommand \enquote  [1]{``#1''}%
\providecommand \bibnamefont  [1]{#1}%
\providecommand \bibfnamefont [1]{#1}%
\providecommand \citenamefont [1]{#1}%
\providecommand \href@noop [0]{\@secondoftwo}%
\providecommand \href [0]{\begingroup \@sanitize@url \@href}%
\providecommand \@href[1]{\@@startlink{#1}\@@href}%
\providecommand \@@href[1]{\endgroup#1\@@endlink}%
\providecommand \@sanitize@url [0]{\catcode `\\12\catcode `\$12\catcode
  `\&12\catcode `\#12\catcode `\^12\catcode `\_12\catcode `\%12\relax}%
\providecommand \@@startlink[1]{}%
\providecommand \@@endlink[0]{}%
\providecommand \url  [0]{\begingroup\@sanitize@url \@url }%
\providecommand \@url [1]{\endgroup\@href {#1}{\urlprefix }}%
\providecommand \urlprefix  [0]{URL }%
\providecommand \Eprint [0]{\href }%
\providecommand \doibase [0]{http://dx.doi.org/}%
\providecommand \selectlanguage [0]{\@gobble}%
\providecommand \bibinfo  [0]{\@secondoftwo}%
\providecommand \bibfield  [0]{\@secondoftwo}%
\providecommand \translation [1]{[#1]}%
\providecommand \BibitemOpen [0]{}%
\providecommand \bibitemStop [0]{}%
\providecommand \bibitemNoStop [0]{.\EOS\space}%
\providecommand \EOS [0]{\spacefactor3000\relax}%
\providecommand \BibitemShut  [1]{\csname bibitem#1\endcsname}%
\let\auto@bib@innerbib\@empty
\bibitem [{\citenamefont {Woodford}\ and\ \citenamefont
  {Livermore}(2009)}]{Woodford09}%
  \BibitemOpen
  \bibfield  {author} {\bibinfo {author} {\bibfnamefont {N.}~\bibnamefont
  {Woodford}}\ and\ \bibinfo {author} {\bibfnamefont {D.~M.}\ \bibnamefont
  {Livermore}},\ }\href@noop {} {\bibfield  {journal} {\bibinfo  {journal}
  {Journal of Infection}\ }\textbf {\bibinfo {volume} {59}},\ \bibinfo {pages}
  {S4} (\bibinfo {year} {2009})}\BibitemShut {NoStop}%
\bibitem [{\citenamefont {Andersson}\ and\ \citenamefont
  {Hughes}(2010)}]{Andersson10}%
  \BibitemOpen
  \bibfield  {author} {\bibinfo {author} {\bibfnamefont {D.~I.}\ \bibnamefont
  {Andersson}}\ and\ \bibinfo {author} {\bibfnamefont {D.}~\bibnamefont
  {Hughes}},\ }\href@noop {} {\bibfield  {journal} {\bibinfo  {journal} {Nature
  Reviews Microbiology}\ }\textbf {\bibinfo {volume} {8}},\ \bibinfo {pages}
  {260} (\bibinfo {year} {2010})}\BibitemShut {NoStop}%
\bibitem [{\citenamefont {Charles}\ and\ \citenamefont
  {Grayson}(2004)}]{Charles04}%
  \BibitemOpen
  \bibfield  {author} {\bibinfo {author} {\bibfnamefont {P.~G.}\ \bibnamefont
  {Charles}}\ and\ \bibinfo {author} {\bibfnamefont {M.~L.}\ \bibnamefont
  {Grayson}},\ }\href@noop {} {\bibfield  {journal} {\bibinfo  {journal}
  {Medical Journal of Australia}\ }\textbf {\bibinfo {volume} {181}},\ \bibinfo
  {pages} {549} (\bibinfo {year} {2004})}\BibitemShut {NoStop}%
\bibitem [{WHO(2015)}]{WHO15}%
  \BibitemOpen
  \href@noop {} {\emph {\bibinfo {title} {Global Action Plan on Antibiotic
  Resistance}}},\ \bibinfo {type} {Tech. Rep.}\ (\bibinfo  {institution} {World
  Health Organisation/UNO},\ \bibinfo {year} {2015})\BibitemShut {NoStop}%
\bibitem [{\citenamefont {Ng}\ \emph {et~al.}(2010)\citenamefont {Ng},
  \citenamefont {Chatenay}, \citenamefont {Robert},\ and\ \citenamefont
  {Poirier}}]{Ng10}%
  \BibitemOpen
  \bibfield  {author} {\bibinfo {author} {\bibfnamefont {J.~W.}\ \bibnamefont
  {Ng}}, \bibinfo {author} {\bibfnamefont {D.}~\bibnamefont {Chatenay}},
  \bibinfo {author} {\bibfnamefont {J.}~\bibnamefont {Robert}}, \ and\ \bibinfo
  {author} {\bibfnamefont {M.~G.}\ \bibnamefont {Poirier}},\ }\href@noop {}
  {\bibfield  {journal} {\bibinfo  {journal} {Phys. Rev. E}\ }\textbf {\bibinfo
  {volume} {81}},\ \bibinfo {pages} {011909} (\bibinfo {year}
  {2010})}\BibitemShut {NoStop}%
\bibitem [{\citenamefont {San~Millan}\ \emph {et~al.}(2015)\citenamefont
  {San~Millan}, \citenamefont {Toll-Riera}, \citenamefont {Qi},\ and\
  \citenamefont {MacLean}}]{Millan15}%
  \BibitemOpen
  \bibfield  {author} {\bibinfo {author} {\bibfnamefont {A.}~\bibnamefont
  {San~Millan}}, \bibinfo {author} {\bibfnamefont {M.}~\bibnamefont
  {Toll-Riera}}, \bibinfo {author} {\bibfnamefont {Q.}~\bibnamefont {Qi}}, \
  and\ \bibinfo {author} {\bibfnamefont {R.~C.}\ \bibnamefont {MacLean}},\
  }\href@noop {} {\bibfield  {journal} {\bibinfo  {journal} {Nature
  Communications}\ }\textbf {\bibinfo {volume} {6}} (\bibinfo {year}
  {2015})}\BibitemShut {NoStop}%
\bibitem [{\citenamefont {Baltrus}(2013)}]{Baltrus13}%
  \BibitemOpen
  \bibfield  {author} {\bibinfo {author} {\bibfnamefont {D.~A.}\ \bibnamefont
  {Baltrus}},\ }\href@noop {} {\bibfield  {journal} {\bibinfo  {journal}
  {Trends in Ecology \& Evolution}\ }\textbf {\bibinfo {volume} {28}},\
  \bibinfo {pages} {489} (\bibinfo {year} {2013})}\BibitemShut {NoStop}%
\bibitem [{\citenamefont {Barbosa}\ and\ \citenamefont
  {Levy}(2000)}]{Barbosa00}%
  \BibitemOpen
  \bibfield  {author} {\bibinfo {author} {\bibfnamefont {T.~M.}\ \bibnamefont
  {Barbosa}}\ and\ \bibinfo {author} {\bibfnamefont {S.~B.}\ \bibnamefont
  {Levy}},\ }\href@noop {} {\bibfield  {journal} {\bibinfo  {journal} {Drug
  Resistance Updates}\ }\textbf {\bibinfo {volume} {3}},\ \bibinfo {pages}
  {303} (\bibinfo {year} {2000})}\BibitemShut {NoStop}%
\bibitem [{\citenamefont {Austin}\ \emph {et~al.}(1999)\citenamefont {Austin},
  \citenamefont {Kristinsson},\ and\ \citenamefont {Anderson}}]{Austin99}%
  \BibitemOpen
  \bibfield  {author} {\bibinfo {author} {\bibfnamefont {D.~J.}\ \bibnamefont
  {Austin}}, \bibinfo {author} {\bibfnamefont {K.~G.}\ \bibnamefont
  {Kristinsson}}, \ and\ \bibinfo {author} {\bibfnamefont {R.~M.}\ \bibnamefont
  {Anderson}},\ }\href@noop {} {\bibfield  {journal} {\bibinfo  {journal}
  {Proceedings of the National Academy of Sciences}\ }\textbf {\bibinfo
  {volume} {96}},\ \bibinfo {pages} {1152} (\bibinfo {year}
  {1999})}\BibitemShut {NoStop}%
\bibitem [{\citenamefont {Tan}\ \emph {et~al.}(2011)\citenamefont {Tan},
  \citenamefont {Serene}, \citenamefont {Chao},\ and\ \citenamefont
  {Gore}}]{Tan11}%
  \BibitemOpen
  \bibfield  {author} {\bibinfo {author} {\bibfnamefont {L.}~\bibnamefont
  {Tan}}, \bibinfo {author} {\bibfnamefont {S.}~\bibnamefont {Serene}},
  \bibinfo {author} {\bibfnamefont {H.~X.}\ \bibnamefont {Chao}}, \ and\
  \bibinfo {author} {\bibfnamefont {J.}~\bibnamefont {Gore}},\ }\href@noop {}
  {\bibfield  {journal} {\bibinfo  {journal} {Phys. Rev. Lett.}\ }\textbf
  {\bibinfo {volume} {106}},\ \bibinfo {pages} {198102} (\bibinfo {year}
  {2011})}\BibitemShut {NoStop}%
\bibitem [{\citenamefont {Bonhoeffer}\ \emph {et~al.}(1997)\citenamefont
  {Bonhoeffer}, \citenamefont {Lipsitch},\ and\ \citenamefont
  {Levin}}]{Bonhoeffer97}%
  \BibitemOpen
  \bibfield  {author} {\bibinfo {author} {\bibfnamefont {S.}~\bibnamefont
  {Bonhoeffer}}, \bibinfo {author} {\bibfnamefont {M.}~\bibnamefont
  {Lipsitch}}, \ and\ \bibinfo {author} {\bibfnamefont {B.~R.}\ \bibnamefont
  {Levin}},\ }\href@noop {} {\bibfield  {journal} {\bibinfo  {journal}
  {Proceedings of the National Academy of Sciences}\ }\textbf {\bibinfo
  {volume} {94}},\ \bibinfo {pages} {12106} (\bibinfo {year}
  {1997})}\BibitemShut {NoStop}%
\bibitem [{\citenamefont {Alexander}\ \emph {et~al.}(2007)\citenamefont
  {Alexander}, \citenamefont {Bowman}, \citenamefont {Feng}, \citenamefont
  {Gardam}, \citenamefont {Moghadas}, \citenamefont {R{\"o}st}, \citenamefont
  {Wu},\ and\ \citenamefont {Yan}}]{Alexander07}%
  \BibitemOpen
  \bibfield  {author} {\bibinfo {author} {\bibfnamefont {M.~E.}\ \bibnamefont
  {Alexander}}, \bibinfo {author} {\bibfnamefont {C.~S.}\ \bibnamefont
  {Bowman}}, \bibinfo {author} {\bibfnamefont {Z.}~\bibnamefont {Feng}},
  \bibinfo {author} {\bibfnamefont {M.}~\bibnamefont {Gardam}}, \bibinfo
  {author} {\bibfnamefont {S.~M.}\ \bibnamefont {Moghadas}}, \bibinfo {author}
  {\bibfnamefont {G.}~\bibnamefont {R{\"o}st}}, \bibinfo {author}
  {\bibfnamefont {J.}~\bibnamefont {Wu}}, \ and\ \bibinfo {author}
  {\bibfnamefont {P.}~\bibnamefont {Yan}},\ }\href@noop {} {\bibfield
  {journal} {\bibinfo  {journal} {Proceedings of the Royal Society of London B:
  Biological Sciences}\ }\textbf {\bibinfo {volume} {274}},\ \bibinfo {pages}
  {1675} (\bibinfo {year} {2007})}\BibitemShut {NoStop}%
\bibitem [{\citenamefont {D'Agata}\ \emph {et~al.}(2008)\citenamefont
  {D'Agata}, \citenamefont {Dupont-Rouzeyrol}, \citenamefont {Magal},
  \citenamefont {Olivier},\ and\ \citenamefont {Ruan}}]{DAgata08}%
  \BibitemOpen
  \bibfield  {author} {\bibinfo {author} {\bibfnamefont {E.~M.}\ \bibnamefont
  {D'Agata}}, \bibinfo {author} {\bibfnamefont {M.}~\bibnamefont
  {Dupont-Rouzeyrol}}, \bibinfo {author} {\bibfnamefont {P.}~\bibnamefont
  {Magal}}, \bibinfo {author} {\bibfnamefont {D.}~\bibnamefont {Olivier}}, \
  and\ \bibinfo {author} {\bibfnamefont {S.}~\bibnamefont {Ruan}},\ }\href@noop
  {} {\bibfield  {journal} {\bibinfo  {journal} {PLoS One}\ }\textbf {\bibinfo
  {volume} {3}},\ \bibinfo {pages} {e4036} (\bibinfo {year}
  {2008})}\BibitemShut {NoStop}%
\bibitem [{\citenamefont {Obolski}\ and\ \citenamefont
  {Hadany}(2012)}]{Obolski12}%
  \BibitemOpen
  \bibfield  {author} {\bibinfo {author} {\bibfnamefont {U.}~\bibnamefont
  {Obolski}}\ and\ \bibinfo {author} {\bibfnamefont {L.}~\bibnamefont
  {Hadany}},\ }\href@noop {} {\bibfield  {journal} {\bibinfo  {journal} {BMC
  Medicine}\ }\textbf {\bibinfo {volume} {10}},\ \bibinfo {pages} {89}
  (\bibinfo {year} {2012})}\BibitemShut {NoStop}%
\bibitem [{\citenamefont {Ternent}\ \emph {et~al.}(2015)\citenamefont
  {Ternent}, \citenamefont {Dyson}, \citenamefont {Krachler},\ and\
  \citenamefont {Jabbari}}]{Ternent15}%
  \BibitemOpen
  \bibfield  {author} {\bibinfo {author} {\bibfnamefont {L.}~\bibnamefont
  {Ternent}}, \bibinfo {author} {\bibfnamefont {R.~J.}\ \bibnamefont {Dyson}},
  \bibinfo {author} {\bibfnamefont {A.-M.}\ \bibnamefont {Krachler}}, \ and\
  \bibinfo {author} {\bibfnamefont {S.}~\bibnamefont {Jabbari}},\ }\href@noop
  {} {\bibfield  {journal} {\bibinfo  {journal} {Journal of Theoretical
  Biology}\ }\textbf {\bibinfo {volume} {372}},\ \bibinfo {pages} {1} (\bibinfo
  {year} {2015})}\BibitemShut {NoStop}%
\bibitem [{\citenamefont {Patra}\ and\ \citenamefont {Klumpp}(2014)}]{Patra14}%
  \BibitemOpen
  \bibfield  {author} {\bibinfo {author} {\bibfnamefont {P.}~\bibnamefont
  {Patra}}\ and\ \bibinfo {author} {\bibfnamefont {S.}~\bibnamefont {Klumpp}},\
  }\href@noop {} {\bibfield  {journal} {\bibinfo  {journal} {Phys. Rev. E}\
  }\textbf {\bibinfo {volume} {89}},\ \bibinfo {pages} {030702} (\bibinfo
  {year} {2014})}\BibitemShut {NoStop}%
\bibitem [{\citenamefont {Lambert}\ and\ \citenamefont
  {Kussell}(2015)}]{Lambert15}%
  \BibitemOpen
  \bibfield  {author} {\bibinfo {author} {\bibfnamefont {G.}~\bibnamefont
  {Lambert}}\ and\ \bibinfo {author} {\bibfnamefont {E.}~\bibnamefont
  {Kussell}},\ }\href@noop {} {\bibfield  {journal} {\bibinfo  {journal} {Phys.
  Rev. X}\ }\textbf {\bibinfo {volume} {5}},\ \bibinfo {pages} {011016}
  (\bibinfo {year} {2015})}\BibitemShut {NoStop}%
\bibitem [{\citenamefont {Bittihn}\ \emph {et~al.}(2017)\citenamefont
  {Bittihn}, \citenamefont {Hasty},\ and\ \citenamefont
  {Tsimring}}]{Bittihn17}%
  \BibitemOpen
  \bibfield  {author} {\bibinfo {author} {\bibfnamefont {P.}~\bibnamefont
  {Bittihn}}, \bibinfo {author} {\bibfnamefont {J.}~\bibnamefont {Hasty}}, \
  and\ \bibinfo {author} {\bibfnamefont {L.~S.}\ \bibnamefont {Tsimring}},\
  }\href@noop {} {\bibfield  {journal} {\bibinfo  {journal} {Phys. Rev. Lett.}\
  }\textbf {\bibinfo {volume} {118}},\ \bibinfo {pages} {028102} (\bibinfo
  {year} {2017})}\BibitemShut {NoStop}%
\bibitem [{\citenamefont {Hermsen}\ and\ \citenamefont
  {Hwa}(2010)}]{Hermsen10}%
  \BibitemOpen
  \bibfield  {author} {\bibinfo {author} {\bibfnamefont {R.}~\bibnamefont
  {Hermsen}}\ and\ \bibinfo {author} {\bibfnamefont {T.}~\bibnamefont {Hwa}},\
  }\href@noop {} {\bibfield  {journal} {\bibinfo  {journal} {Phys. Rev. Lett.}\
  }\textbf {\bibinfo {volume} {105}},\ \bibinfo {pages} {248104} (\bibinfo
  {year} {2010})}\BibitemShut {NoStop}%
\bibitem [{\citenamefont {Greulich}\ \emph {et~al.}(2012)\citenamefont
  {Greulich}, \citenamefont {Waclaw},\ and\ \citenamefont
  {Allen}}]{Greulich12}%
  \BibitemOpen
  \bibfield  {author} {\bibinfo {author} {\bibfnamefont {P.}~\bibnamefont
  {Greulich}}, \bibinfo {author} {\bibfnamefont {B.}~\bibnamefont {Waclaw}}, \
  and\ \bibinfo {author} {\bibfnamefont {R.~J.}\ \bibnamefont {Allen}},\
  }\href@noop {} {\bibfield  {journal} {\bibinfo  {journal} {Phys. Rev. Lett.}\
  }\textbf {\bibinfo {volume} {109}},\ \bibinfo {pages} {088101} (\bibinfo
  {year} {2012})}\BibitemShut {NoStop}%
\bibitem [{\citenamefont {Tikhonov}\ and\ \citenamefont
  {Monasson}(2017)}]{Tikhonov17}%
  \BibitemOpen
  \bibfield  {author} {\bibinfo {author} {\bibfnamefont {M.}~\bibnamefont
  {Tikhonov}}\ and\ \bibinfo {author} {\bibfnamefont {R.}~\bibnamefont
  {Monasson}},\ }\href@noop {} {\bibfield  {journal} {\bibinfo  {journal}
  {Phys. Rev. Lett.}\ }\textbf {\bibinfo {volume} {118}},\ \bibinfo {pages}
  {048103} (\bibinfo {year} {2017})}\BibitemShut {NoStop}%
\bibitem [{\citenamefont {Davey}\ \emph {et~al.}(2015)\citenamefont {Davey},
  \citenamefont {Wilcox}, \citenamefont {Irving}, \citenamefont {Thwaites}
  \emph {et~al.}}]{Davey15}%
  \BibitemOpen
  \bibfield  {author} {\bibinfo {author} {\bibfnamefont {P.~G.}\ \bibnamefont
  {Davey}}, \bibinfo {author} {\bibfnamefont {M.~H.}\ \bibnamefont {Wilcox}},
  \bibinfo {author} {\bibfnamefont {W.~L.}\ \bibnamefont {Irving}}, \bibinfo
  {author} {\bibfnamefont {G.}~\bibnamefont {Thwaites}},  \emph {et~al.},\
  }\href@noop {} {\emph {\bibinfo {title} {Antimicrobial chemotherapy}}}\
  (\bibinfo  {publisher} {Oxford University Press, USA},\ \bibinfo {year}
  {2015})\BibitemShut {NoStop}%
\bibitem [{\citenamefont {French}(2006)}]{French06}%
  \BibitemOpen
  \bibfield  {author} {\bibinfo {author} {\bibfnamefont {G.}~\bibnamefont
  {French}},\ }\href@noop {} {\bibfield  {journal} {\bibinfo  {journal}
  {Journal of Antimicrobial Chemotherapy}\ }\textbf {\bibinfo {volume} {58}},\
  \bibinfo {pages} {1107} (\bibinfo {year} {2006})}\BibitemShut {NoStop}%
\bibitem [{\citenamefont {Stadler}\ and\ \citenamefont
  {Stadler}(2006)}]{Stadler06}%
  \BibitemOpen
  \bibfield  {author} {\bibinfo {author} {\bibfnamefont {P.~F.}\ \bibnamefont
  {Stadler}}\ and\ \bibinfo {author} {\bibfnamefont {B.~M.}\ \bibnamefont
  {Stadler}},\ }\href@noop {} {\bibfield  {journal} {\bibinfo  {journal}
  {Biological Theory}\ }\textbf {\bibinfo {volume} {1}},\ \bibinfo {pages}
  {268} (\bibinfo {year} {2006})}\BibitemShut {NoStop}%
\bibitem [{\citenamefont {Cortini}\ \emph {et~al.}(2016)\citenamefont
  {Cortini}, \citenamefont {Barbi}, \citenamefont {Car\'e}, \citenamefont
  {Lavelle}, \citenamefont {Lesne}, \citenamefont {Mozziconacci},\ and\
  \citenamefont {Victor}}]{Cortini16}%
  \BibitemOpen
  \bibfield  {author} {\bibinfo {author} {\bibfnamefont {R.}~\bibnamefont
  {Cortini}}, \bibinfo {author} {\bibfnamefont {M.}~\bibnamefont {Barbi}},
  \bibinfo {author} {\bibfnamefont {B.~R.}\ \bibnamefont {Car\'e}}, \bibinfo
  {author} {\bibfnamefont {C.}~\bibnamefont {Lavelle}}, \bibinfo {author}
  {\bibfnamefont {A.}~\bibnamefont {Lesne}}, \bibinfo {author} {\bibfnamefont
  {J.}~\bibnamefont {Mozziconacci}}, \ and\ \bibinfo {author} {\bibfnamefont
  {J.-M.}\ \bibnamefont {Victor}},\ }\href@noop {} {\bibfield  {journal}
  {\bibinfo  {journal} {Rev. Mod. Phys.}\ }\textbf {\bibinfo {volume} {88}},\
  \bibinfo {pages} {025002} (\bibinfo {year} {2016})}\BibitemShut {NoStop}%
\bibitem [{\citenamefont {Rosenblatt}(1958)}]{Rosenblatt58}%
  \BibitemOpen
  \bibfield  {author} {\bibinfo {author} {\bibfnamefont {F.}~\bibnamefont
  {Rosenblatt}},\ }\href@noop {} {\bibfield  {journal} {\bibinfo  {journal}
  {Psychological Review}\ }\textbf {\bibinfo {volume} {65}},\ \bibinfo {pages}
  {386} (\bibinfo {year} {1958})}\BibitemShut {NoStop}%
\bibitem [{\citenamefont {Engel}\ and\ \citenamefont {van~den
  Broeck}(2001)}]{Engel01}%
  \BibitemOpen
  \bibfield  {author} {\bibinfo {author} {\bibfnamefont {A.}~\bibnamefont
  {Engel}}\ and\ \bibinfo {author} {\bibfnamefont {C.}~\bibnamefont {van~den
  Broeck}},\ }\href@noop {} {\emph {\bibinfo {title} {Statistical Mechanics of
  Learning}}}\ (\bibinfo  {publisher} {Cambridge University Press},\ \bibinfo
  {year} {2001})\BibitemShut {NoStop}%
\bibitem [{\citenamefont {Cybenko}(1989)}]{Cybenko89}%
  \BibitemOpen
  \bibfield  {author} {\bibinfo {author} {\bibfnamefont {G.}~\bibnamefont
  {Cybenko}},\ }\href@noop {} {\bibfield  {journal} {\bibinfo  {journal}
  {Mathematics of Control, Signals and Systems}\ }\textbf {\bibinfo {volume}
  {2}},\ \bibinfo {pages} {303} (\bibinfo {year} {1989})}\BibitemShut {NoStop}%
\bibitem [{\citenamefont {Silver}\ \emph {et~al.}(2016)\citenamefont {Silver},
  \citenamefont {Huang}, \citenamefont {Maddison}, \citenamefont {Guez},
  \citenamefont {Sifre}, \citenamefont {Van Den~Driessche}, \citenamefont
  {Schrittwieser}, \citenamefont {Antonoglou}, \citenamefont {Panneershelvam},
  \citenamefont {Lanctot} \emph {et~al.}}]{Silver16}%
  \BibitemOpen
  \bibfield  {author} {\bibinfo {author} {\bibfnamefont {D.}~\bibnamefont
  {Silver}}, \bibinfo {author} {\bibfnamefont {A.}~\bibnamefont {Huang}},
  \bibinfo {author} {\bibfnamefont {C.~J.}\ \bibnamefont {Maddison}}, \bibinfo
  {author} {\bibfnamefont {A.}~\bibnamefont {Guez}}, \bibinfo {author}
  {\bibfnamefont {L.}~\bibnamefont {Sifre}}, \bibinfo {author} {\bibfnamefont
  {G.}~\bibnamefont {Van Den~Driessche}}, \bibinfo {author} {\bibfnamefont
  {J.}~\bibnamefont {Schrittwieser}}, \bibinfo {author} {\bibfnamefont
  {I.}~\bibnamefont {Antonoglou}}, \bibinfo {author} {\bibfnamefont
  {V.}~\bibnamefont {Panneershelvam}}, \bibinfo {author} {\bibfnamefont
  {M.}~\bibnamefont {Lanctot}},  \emph {et~al.},\ }\href@noop {} {\bibfield
  {journal} {\bibinfo  {journal} {Nature}\ }\textbf {\bibinfo {volume} {529}},\
  \bibinfo {pages} {484} (\bibinfo {year} {2016})}\BibitemShut {NoStop}%
\bibitem [{\citenamefont {Park}\ and\ \citenamefont {Deem}(2007)}]{Park07}%
  \BibitemOpen
  \bibfield  {author} {\bibinfo {author} {\bibfnamefont {J.-M.}\ \bibnamefont
  {Park}}\ and\ \bibinfo {author} {\bibfnamefont {M.~W.}\ \bibnamefont
  {Deem}},\ }\href@noop {} {\bibfield  {journal} {\bibinfo  {journal} {Phys.
  Rev. Lett.}\ }\textbf {\bibinfo {volume} {98}},\ \bibinfo {pages} {058101}
  (\bibinfo {year} {2007})}\BibitemShut {NoStop}%
\end{thebibliography}%

\end{document}